Center of mass velocity from Mass Polariton solution to Abraham-Minkowski controversy.

J. P. McClymer [*]

Department of Physics and Astronomy, University of Maine, 5709 Bennett, Orono, ME 04469-5709, United States

**ABSTRACT**

A one-hundred-year old controversy, the Abraham-Minkowski controversy, concerns two widely disparate predictions for the momentum of light in glass. In the Abraham case the photon momentum is inversely proportional to the index of refraction while the photon momentum predicted by Minkowski formulation is proportional to the index of refraction. In spite of significant theoretical and experimental effort neither formulation could convincingly be shown to be correct. A new model, called the Mass Polariton model, MP, was recently introduced that resolves the contradictions in the Abraham and Minkowski formulations and experimental results. Both the MP and Minkowski model predict the same momentum change for a photon entering glass from a vacuum, yet the Minkowski model allows violation of the center of mass theorem. This paper shows that the MP model, in spite of predicting the same momentum change as the Minkowski model, obeys the center of mass theorem.

Keywords: Electrodynamics; Abraham-Minkowski controversy, conservation of momentum

## 1. Introduction

There are a surprising number of issues in physics where there is a substantial literature, and sometimes controversy, that remain unknown to large numbers of the community. The Abraham-

[*] Corresponding author.
E-mail address: McClymer@maine.edu

Minkowski controversy is an example of just such an issue. The 21st century has brought renewed interest with well-publicized "solutions"[1,2] to the controversy, culminating in a theoretical model that appears to resolve the controversy using fundamental physical principles.

The problem is simply stated; what is the momentum of a photon in glass? For our purposes the glass is a transparent, non-scattering, non-dispersive dielectric with real index of refraction, n. The two leading models were introduced in the early part of the 20th century. Abraham[3] and Minkowski[4,5] provide differing definitions for the electromagnetic momentum-energy tensor of light in a dielectric. These differing tensors result in different predictions for the momentum density;

$$\vec{g}_{Abraham} = \vec{E} \times \frac{\vec{H}}{c^2}, \vec{g}_{Minkowski} = \vec{D} \times \vec{B}. \tag{1}$$

These differing momentum densities then give contradictory predictions for the momentum of a photon in a dielectric of index n;

$$\left|\vec{p}_{\gamma Abraham}\right| = \frac{\hbar\omega}{nc}, \left|\vec{p}_{\gamma Minkowski}\right| = \frac{n\hbar\omega}{c}. \tag{2}$$

While both momenta equations agree in a vacuum, they predict very different behavior in a material, behavior so different that it would seem trivial to experimentally determine which, if either, is valid. The experiments however are often subtle, with some results favoring the Abraham, and some a Minkowski, interpretation. Significant theoretical effort has been made over the last century to resolve this question, yet the question remains into the 21st century.

I will not review the extensive literature on this subject, instead suggesting some recent review articles[6,7,8] which review the theoretical approaches and the contradictory appearing experimental results through 2010. Many additional papers appeared after this time, some responding to the new results in 2008 and theoretical model in 2010 given in reference 1 and 2.

Both the Abraham and Minkowski theoretical underpinnings appear extremely sound and inviolable. A fundamental argument for the Abraham model is that of the idea of center of mass and the well-established center of mass theorem which states that the center of mass of an isolated system moves at a uniform velocity, a corollary of momentum conservation. Momentum conservation prevents existence of a reactionless drive; an isolated system that changes its momentum. While the Abraham model prevents a reactionless drive, the Minkowski model allows it[9]. Normally, a model violating such a fundamental tenant would be easily discarded but the evidence in favor of Minkowski is just as substantial as that supporting the Abraham model. For instance, the Minkowski model is easily argued from the de Broglie hypothesis and the known reduction of wavelength in a dielectric;

$$\left|\vec{p}_{\gamma Minkowski}\right| = \frac{2\pi\hbar}{\lambda} = \frac{2\pi n\hbar}{\lambda_0} = \frac{n\hbar\omega}{c} \quad , \tag{3}$$

with $\lambda = \frac{\lambda_0}{n}$ and $\lambda_0$ being the wavelength in vacuum.

I describe a simple thought experiment to demonstrate that the Abraham momentum, as it must, does not allow a reactionless drive while the Minkowski model for momentum does predict non-uniform movement of the center of mass of an isolated system. I then use the Mass Polariton[10] model (MP) to show, even though its key result, the momentum of the "photon" is the same as the Minkowski value, does not lead to non-uniform motion of the center of mass. The Mass Polariton model has since been extended[11] with further work.

## 2. Thought experiment

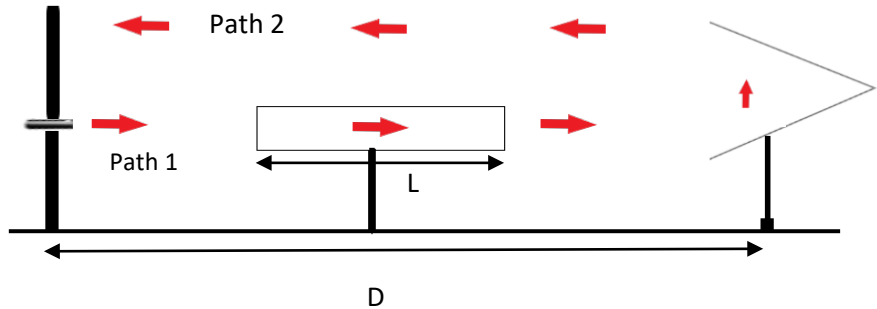

*Figure 1* A photon, represented by arrows, travels from a laser, in vacuum, through a dielectric (black rectangle) of length L, back into vacuum, then into a retroreflector which returns the photon to the left where it is absorbed. The dielectric, the laser, and reflector as well as all mounts are mounted to the isolated platform of length D.

Figure 1 shows a platform of length D with a laser at the left end and a retroreflector at the other end that reflects the photon back to the left where it is absorbed. A dielectric material, represented by the rectangular block which I refer to as "glass", partially fills the space as shown. The glass, of length L, is non-reflecting with an anti-reflection coating (not shown), non-absorbing, non-dispersive with a real refractive index n and mass M. These are the same material attributes assumed in the MP paper. The platform, which is isolated from the rest of the universe, has total mass $M_P$ which is the mass of all the components including the laser, dielectric, retroreflector and mounts. This platform mass does not include the mass-energy of light emitted from the laser. The platform is initially at rest in the lab frame.

The closed path (the photon could be reflected downward and absorbed arbitrarily close to the source) the light takes can be considered as consisting of two paths. In path 1 (photon traveling left to right), the photon travels a distance in vacuum, enters and travels through a dielectric, before exiting the dielectric and reaching the retroreflector. In path 2 the photon travels only through vacuum back to the left side where it is absorbed. As we follow the photon from its start on the left, along path 1 and return through path 2 we expect no net-displacement of the platform, as predicted by the center of mass theorem.

Let's analyze the thought experiment one step at a time and see why it could *potentially* allow non-uniform motion of the center of mass; i.e. a reactionless drive. Using the directions in Figure 1, the launch of the photon imparts a momentum to the platform to the left. The photon then encounters the glass, travels through the glass for some time at a lower speed, before exiting the glass into vacuum. The momentum change of the photon as it first enters, then exist the glass gives a net momentum to the platform, $\vec{p}_{Glass} = \vec{p}_{\gamma_{vacuum}} - \vec{p}_{\gamma G}$, with the last term is the momentum of the photon in the glass. Depending on the value of the photon momentum in the glass, the momentum of the glass could be positive, to the right, (Abraham) or negative, to the left, (Minkowski).

Once the vacuum photon strikes the right side of the platform the initial platform momentum from the photon launch is cancelled out. Yet, due to the change in momentum of the photon as it enters the glass, the platform has a non-zero displacement. The value of this displacement depends on the photon momentum in the glass.

When the photon reflects off the right side of the platform, the platform recoils to the right and comes to a stop when the photon reaches the left side of the platform and is absorbed. This return trip occurs in vacuum. The question then is if the net displacement is zero or not.

The above description is quantified as follows; I suppress vector signs and define motion to the left as negative and motion to the right as positive. The laser, emitting light, gives momentum, -p₁, to the platform for a duration, $t = \frac{(D-L)}{c} + \frac{nL}{c}$. While the light is in the glass the platform momentum changes by p_glass, which acts for a time $t = \frac{nL}{c}$. The reflection of the light for the return trip gives the platform momentum +p₁ for time, $t = \frac{D}{c}$.

The total displacement, due to round trip of the light, is

$$d_{net} = \frac{L}{M_P c}\left(n\, p_{glass} + p_1(1-n)\right) = \frac{L}{M_P c}\left(p_1 - n\, p_{\gamma G}\right). \tag{4}$$

The platform momentum due to the photon launch is $p_1 = \frac{\hbar \omega}{c}$. Substituting in for the photon momentum in glass, the Abraham, $p_{\gamma G_A} = \frac{\hbar \omega}{nc}$, and then Minkowski, $p_{\gamma G_M} = \frac{n\hbar \omega}{c}$, gives a zero displacement for the Abraham photon while the Minkowski photon yields a non-zero displacement of $d_{Minkowski} = \left(\frac{L}{M_P c}\right)\left(\frac{\hbar \omega}{c}\right)(1-n^2)$.

As emphasized earlier, this unexpected result is not sufficient to rule out the Minkowski model, although it is an indication of something deeper occurring.

It has long been recognized that neither the Abraham nor Minkowski photon values satisfy the Einstein covariance principle;

$$E^2 - (pc)^2 = (mc^2)^2. \tag{5}$$

Since photons have no mass, no model that considers only the field term can be correct. In addition to a field term, a term with mass, from the stress tensor of the material, must be included. The "splitting" of momentum between the field and material terms is not arbitrary, as proposed in reference 6 and references therein. The MP model posits that the photon/dielectric forms a mass-polariton quasi-particle (MP) that satisfies the covariance principle. The authors point out that their definition of mass-polariton is distinct from the conventional usage of the term as it does not involve a close resonance with internal states of the system.

Henceforth, I adopt the terminology of the MP authors. My equations 5 through 8 are from the MP paper, but with different numbering. I suppress vector signs on all momenta, with positive values meaning momentum to the right and negative values meaning momentum to the left, consistent with Figure 1. The vacuum photon, with energy $E = \hbar\omega$, and momentum $p_{\gamma_{vacuum}} = \frac{\hbar\omega}{c}$, on entering the dielectric creates an MP, which consists of a field component and a mass density wave (MDW), propagating together at speed $v = \frac{c}{n}$ inside the dielectric. In the lab frame, the momentum of the MP (field and matter wave) is

$$p_{MP} = \frac{n\hbar\omega}{c}, \tag{6}$$

same momentum as the Minkowski photon (field term). Since the momentum of the MP is the same as for a Minkowski photon, it could be possible that the MP model allows for a reactionless drive. This will not turn out to be the case. The energy of the MP is

$$E_{MP} = n^2\hbar\omega = \gamma m_0 c^2, \tag{7}$$

where $m_0$ is the effective mass of the MP. Note that the momentum of the MP is greater than the momentum of the photon in vacuum, causing the dielectric to move in the opposite direction to propagation direction of the MP, conserving momentum; which is to the left in Figure 1. *The MP* *has $n^2$ more energy than the vacuum photon.* As energy must be conserved, this energy increase comes from a *mass loss* of the dielectric,

$$\partial m = (n^2 - 1)\frac{\hbar \omega}{c^2}. \tag{8}$$

As pointed out by the MP authors, the reaction mass of the dielectric is reduced to

$$M_R = (M - \partial m), \tag{9}$$

where M is the mass of the dielectric before the photon enters, and after it exits, the dielectric. This "mass loss" is not truly lost, it becomes associated instead with the MP quasiparticle moving through the dielectric.

With these equations the MP model allows calculation of the velocity of the center of mass directly.

## 5. Velocity of Center of Energy

The velocity of the center of energy is given by

$$V_{c.m.} = \frac{\sum_i E_i v_i}{\sum_i E_i}. \tag{9}$$

The numerator is proportional to the total momentum of the system while the denominator is its total energy.

Working in the lab frame with the platform initially at rest, when the laser first emits the photon, the platform recoils to the left as the photon travels to the right in vacuum. The velocity of the center of energy is then

$$V_{c.m.} = \frac{\sum_i E_i v_i}{\sum_i E_i} = \frac{\hbar\omega c + M_P c^2 V_{R1}}{\hbar\omega + M_P c^2}. \qquad (10)$$

The recoil velocity of the platform is $V_{R1} = \frac{-\hbar\omega}{M_P c}$ when the laser emits the photon. Putting this velocity into equation 10 shows that the center of energy velocity is unchanged and remains 0, as expected. Note that treating the platform as a single mass with a well-defined velocity, even though the photon travels faster than the impulse giving rise to the velocity, correctly describes the system. The success of this approach is because the numerator is proportional to the momentum in the system.

The situation becomes more complicated when the photon enters the dielectric and creates an MP. The center of energy velocity is then

$$V_{c.m.} = \frac{\sum_i E_i v_i}{\sum_i E_i} = \frac{\gamma m_0 c^2 \left(\frac{c}{n}\right) + M_P c^2 V_{R1} + (M_P - \partial m) c^2 (V_R)}{\gamma m_0 c^2 + (M_P - \partial m) c^2}. \qquad (11)$$

In the above equation, the energy of the MP is $\gamma m_0 c^2$ with velocity c/n. The next term accounts for the momentum of the platform due to the launch of the photon. The final term is due to the reduced mass of the glass and its recoil velocity due to the photon transitioning into an MP. I assume the impulse from the initial photon launch does not reach the glass until after the MP exits the glass.

It is straight forward to calculate the first two terms in the numerator, with the recoil velocity of the platform, $V_{R1}$, defined above. The velocity of the system, $V_R$, when the photon is inside the glass is not so clear. Since there is no such thing as a rigid structure, this velocity is time and position dependent. In Equation 11 $V_R$ represents an average velocity. We can avoid the difficulties inherent in trying to calculate this quantity by using conservation of momentum. Before the photon was launched the system had 0 momentum. The MP was created with momentum given by $p_{MP} = \frac{n\hbar\omega}{c}$. Therefore the momentum of the rest of the system must be $-\left(\frac{n\hbar\omega}{c}\right)$.

Noting that the energy-velocity product of the last two terms in the numerator can be written as $(Mc^2)V$, we know that these terms sum to $-\left(\frac{n\hbar\omega}{c}\right)c^2 = -n\hbar\omega c$.

Substituting in these values the numerator becomes

$$V_{c.m.} \propto \gamma m_0 c^2 \left(\frac{c}{n}\right) + M_P c^2 V_{R1} + (M_P - \partial m)c^2(V_R) = n^2\hbar\omega\left(\frac{c}{n}\right) - n\hbar\omega c = 0. \qquad (12)$$

.

Hence, we can solve for the numerator without know $V_R$ and the result is the center of mass velocity remains unchanged when the photon is inside the glass.

The MP model, even though it predicts the same momentum for the MP as predicted for the photon in the Minkowski model, satisfies the center of mass theorem.

Once we know the value of the last two terms in the numerator from momentum conservation, we can solve for the unknown average velocity, $V_R$, even if the result is not very illuminating.

$$V_R = \frac{(1-n)c}{\dfrac{M_P}{\hbar\omega} - (n^2-1)}. \tag{13}$$

In conclusion, the MP model, even though it predicts the same momentum transfer as the Minkowski model, does not violate the center of mass theorem.

Acknowledgements. I thank Dean Astumian and Neil Comins for helpful discussions and Steve Pollaine for introducing me to the MP model.